\documentclass[10pt,journal,onecolumn]{IEEEtran}
\usepackage{color}
\usepackage{colortbl}
\usepackage{cite}
\usepackage{dcolumn}
\usepackage{amsmath}
\usepackage{amssymb}
\usepackage{graphicx}
\usepackage{epsfig}
\usepackage{changebar}
\usepackage{multirow}
\usepackage{subfigure}
\usepackage{txfonts}
\usepackage{relsize}
\usepackage{verbatim} 
\usepackage{rotating}
\usepackage{lscape}
\usepackage{graphicx}
\usepackage{wrapfig}
\usepackage[top=0.75in, bottom=0.75in, left=0.75in, right=0.75in]{geometry}
\usepackage{setspace}

\newcommand{\aap}{\textit {A\&A}}

\newcommand{\pasa}{\textit {PASA}}

\newcommand\arcmin{\hbox{$^\prime$}}

\newcommand\degr{\hbox{$^\circ$}}
\newcommand*{\TitleFont}{%
      \usefont{\encodingdefault}{\rmdefault}{b}{n}%
      \fontsize{12}{12}%
      \selectfont}
\newcommand{\affil}[1]{$^{\rm #1}$}
\newcounter{inst} 
\newcommand{\inst}[1]{\noindent%
   \refstepcounter{inst}\affil{\Alph{inst}\label{#1}}  % comment if you prefer numeral references
   }
\begin{document}

\title{\TitleFont First Look Murchison Widefield Array observations of Abell 3667}

\author{\textit{\textbf L. Hindson\affil{\ref{VUW}}, M. Johnston-Hollitt\affil{\ref{VUW}}, N. Hurley-Walker\affil{\ref{ICRAR}},
J. Morgan\affil{\ref{ICRAR}},K. Buckley\affil{\ref{VUWCOMP}}, E. Carretti\affil{\ref{CSIRO}}, K.S. Dwarakanath\affil{\ref{RRI}}, M. Bell\affil{\ref{CAASTRO},\ref{USyd}}, 
G. Bernardi\affil{\ref{CfA}}, 
R. Bhat\affil{\ref{ICRAR}}, 
F. Briggs\affil{\ref{CAASTRO},\ref{ANU}}, 
A. A. Deshpande\affil{\ref{RRI}}, 
A. Ewall-Wice\affil{\ref{MIT}}, 
L. Feng\affil{\ref{MIT}}, 
B. Hazelton\affil{\ref{UDub}}, 
D. Jacobs\affil{\ref{ASU}},
D. Kaplan\affil{\ref{UWM}}
N. Kudryavtseva\affil{\ref{ICRAR}}, 
E. Lenc\affil{\ref{CAASTRO},\ref{USyd}}, 
B. McKinley\affil{\ref{CAASTRO},\ref{ANU}}, 
D. Mitchell\affil{\ref{CASS}}, 
B. Pindor\affil{\ref{CAASTRO},\ref{UMelb}}, 
P. Procopio\affil{\ref{CAASTRO},\ref{UMelb}}, 
D. Oberoi\affil{\ref{NCRA}},
A. R. Offringa\affil{\ref{CAASTRO},\ref{ANU}}, 
S. Ord\affil{\ref{ICRAR},\ref{CAASTRO}},
J. Riding\affil{\ref{CAASTRO},\ref{UMelb}},
J. D. Bowman\affil{\ref{ASU}},
R. Cappallo\affil{\ref{Haystack}},
B. Corey\affil{\ref{Haystack}},
D. Emrich\affil{\ref{ICRAR}},
B. M. Gaensler\affil{\ref{CAASTRO},\ref{USyd}},
R. Goeke\affil{\ref{Haystack}},
L. Greenhill\affil{\ref{CfA}},
J. Kasper\affil{\ref{CfA}},
E. Kratzenberg\affil{\ref{Haystack}},
C. Lonsdale\affil{\ref{Haystack}},
M. Lynch\affil{\ref{ICRAR}},
R. McWhirter\affil{\ref{Haystack}},
M. Morales\affil{\ref{UDub}},
E. Morgan\affil{\ref{MIT}},
T. Prabu\affil{\ref{RRI}},
A. Rogers\affil{\ref{Haystack}},
A. Roshi\affil{\ref{NRAO}},
U. Shankar\affil{\ref{RRI}},
K. Srivani\affil{\ref{RRI}},
R. Subrahmanyan\affil{\ref{RRI},\ref{CAASTRO}},
S. Tingay\affil{\ref{ICRAR},\ref{CAASTRO}},
M. Waterson\affil{\ref{ICRAR},\ref{ANU}},
R. Webster\affil{\ref{CAASTRO},\ref{UMelb}},
A. Whitney\affil{\ref{Haystack}},
A. Williams\affil{\ref{ICRAR}},
C. Williams\affil{\ref{MIT}}}
\\

\vspace{0.4cm}
{\small \affil{}\,Email: Luke.Hindson@vuw.ac.nz}\\
{\small \inst{VUW}\,School of Chemical \& Physical Sciences, Victoria University of Wellington, Wellington 6140, New Zealand}\\
{\small \inst{VUWCOMP}\,School of Engineering \& Computer Sciences, Victoria University of Wellington, Wellington 6140, New Zealand}\\
{\small \inst{ICRAR}\,International Centre for Radio Astronomy Research, Curtin University, Bentley, WA 6102, Australia}\\
{\small \inst{CSIRO}\,CSIRO Astronomy and Space Science, Marsfield, NSW 2122, Australia}\\
{\small \inst{RRI}\,Raman Research Institute, Bangalore 560080, India}\\
{\small \inst{ASU}\,School of Earth and Space Exploration, Arizona State University, Tempe, AZ 85287, USA}\\
{\small \inst{NCRA}\,National Centre for Radio Astrophysics, Tata Institute for Fundamental Research, Pune 411007, India}\\
{\small \inst{CAASTRO}\,ARC Centre of Excellence for All-sky Astrophysics (CAASTRO), 44-70 Rosehill Street, Redfern NSW 2016, Sydney, Australia}\\
{\small \inst{USyd}\,Sydney Institute for Astronomy. The University of Sydney, Sydney, Australia}\\
{\small \inst{CfA}\,Harvard-Smithsonian Center for Astrophysics, Cambridge, MA 02138, USA}\\
{\small \inst{ANU}\,Research School of Astronomy and Astrophysics, The Australian National University, Canberra, Australia}\\
{\small \inst{MIT}\,Kavli Institute for Astrophysics and Space Research, Massachusetts Institute of Technology, Cambridge, MA 02139, USA}\\
{\small \inst{UDub}\,Department of Physics, University of Washington, Seattle, WA 98195, USA}\\
{\small \inst{UWM}\,Department of Physics, University of Wisconsin--Milwaukee, Milwaukee, WI 53201, USA}\\
{\small \inst{CASS}\,CSIRO Astronomy and Space Science (CASS), PO Box 76, Epping, NSW 1710, Australia}\\
{\small \inst{UMelb}\,School of Physics, The University of Melbourne, Parkville, VIC 3010, Australia}\\
{\small \inst{Haystack}\,MIT Haystack Observatory, Westford, MA 01886, USA}\\
{\small \inst{NRAO}\,National Radio Astronomy Observatory, Charlottesville and Greenbank, USA}\\
}
\maketitle
\pagenumbering{gobble}

\renewcommand\thesection{\arabic{section}}
\renewcommand\thesubsection{\thesection.\arabic{subsection}}\begin{abstract}
The Murchison Widefield Array (MWA) is a new low frequency interferometric radio telescope, operating in the remote Murchison Radio Observatory in Western Australia. In this paper we present the first MWA observations of the well known radio relics in Abell\,3667 (A3667) between 120 and 226\,MHz. We clearly detect the radio relics in A3667 and present flux estimates and spectral indices for these features. The average spectral index of the north-west (NW) and south-east (SE) relics is $-0.9\pm0.1$ between 120 and 1400\,MHz. We are able to resolve spatial variation in the spectral index of the NW relic from $-1.7$ to $-0.4$, which is consistent with results found at higher frequencies.

\end{abstract}

\section{\textbf {Introduction}}
\singlespacing
Observations of galaxy clusters in the radio regime have revealed low surface brightness and large-scale non-thermal emission up to Mpc-scales associated with some galaxy clusters. These radio sources are known as radio halos and relics and share some observational properties: they have steep synchrotron spectra ($\alpha \sim -1.2$ to $-1.4$) and are characterised by low surface brightness emission ($\sim\mu$\,Jy arcsec$^{-2}$). Radio halos are unpolarised and located towards the centre of the host galaxy cluster, whilst relics are highly polarised and reside on the periphery of the galaxy cluster. 
\newline

The galaxy cluster A3667 is one of the most well studied examples of radio relic emission in the southern sky. The dramatic radio relics on the periphery of the cluster are believed to be generated by shocks in the intra cluster medium generated by two merging clusters. With an angular extent of $\sim 1$\degr\ and redshift of z=0.0556 (distance of 224 Mpc), A3667 extends over $\sim 4$\,Mpc.
\newline

The MWA is one of the Square Kilometre Array precursor telescopes at low frequencies, operating in the regime between 80 and 300\,MHz. Details of the technical design and specifications of the MWA can be found in \cite{Tingay2013} and a review of the scientific goals are presented in \cite{Bowman2013}. The unifying design theme of the MWA emphasises high survey efficiency and surface brightness sensitivity at low frequency. This makes it an ideal instrument with which to carry out large scale untargeted surveys in search for diffuse, steep-spectrum synchrotron emission associated with radio halos and relics. 
\newline

In this paper we present the first MWA observations of A3667 between 120 and 226\,MHz. In section~\ref{Sect:obs}  we present the observations and briefly outline the data reduction process. In section~\ref{Sect:results} we present the results of these observations including images of the radio data, integrated fluxes of emission features and a spectral index map of A3667. 

\section{\textbf {Observations \& data reduction procedures}}\label{Sect:obs}

\begin{wrapfigure}{r}{0.5\textwidth}
    \includegraphics[width=0.55\textwidth,trim = 0 0 0 20]{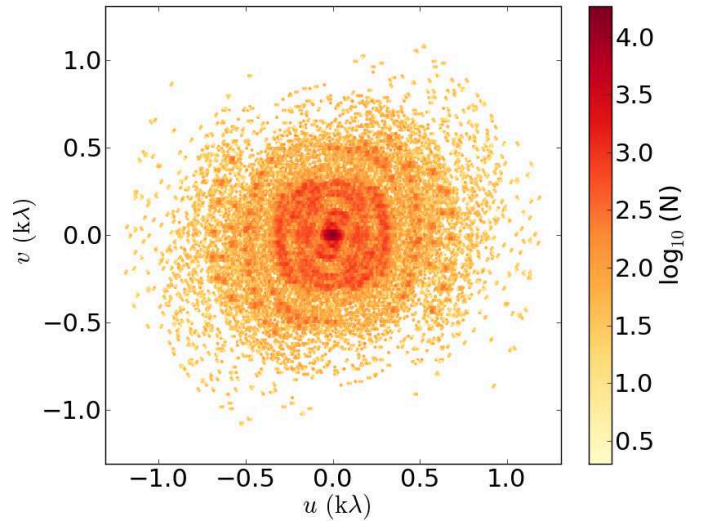}
  \caption{\textbf {Combined {\textit uv} coverage for $\bold {3\times232}$ snapshots centred on 149\,MHz with a bandwidth of 30\,MHz, in units of wavelength. The colour scale indicates the density of $\bold{uv}$ points after applying hexagonal binning.}}
  \label{im:uvcoverage}

\end{wrapfigure}

A3667 was observed using 114 tiles of the full 128 tile MWA array. The observations consist of $3\times232$\,s ``snapshots'' of four 30\,MHz wide bands centred on 120, 149, 180 and 226\,MHz. The minimum and maximum baselines were 7.7 and 2864\,m respectively. This provides sensitivity to structures from $\sim 3$\arcmin\ to several degrees. The combined $uv$ coverage of three snapshots at 149\,MHz, excluding multi frequency synthesis (MFS), can be seen in Fig~.\ref{im:uvcoverage}. The $uv$ coverage at short baselines provides excellent sensitivity to large scale structure, however the low frequency relative to the maximum baseline limits the resolution to $\sim3$\arcmin. 
\newline

\begin{table}
\caption[Observation characteristics]{\textbf{Summary of the observational characteristics of the combined $3\times232$\,s observations.}}
\begin{center}
  \begin{tabular}{cccc}
   \hline Frequency & FOV & Resolution & Sensitivity\\
           (MHz) & degrees ($\degr$) & arc minutes (\arcmin) & (mJy beam$^{-1}$) \\
 	\hline
120	&	15.9	&	$5.3\times3.8$	&	83.3	\\
149	&	13.2	&	$4.2\times3.0$	&	50.0	\\
180	&	11.2	&	$3.6\times2.5$	&	39.5	\\
226	&	9.8		&	$3.0\times2.1$	&	35.1	\\
\hline
\end{tabular}
\end{center}
\label{tab:obs}
\end{table}

The data were processed using both pre-existing and bespoke software written specifically for MWA data processing. The pre-processing pipeline {\sc cotter} was used to perform initial detection and flagging of radio frequency interference using {\sc aoflagger} \cite{Offringa2012} and data averaging to 1\,s and 40\,kHz time and frequency resolution. Gain solutions were determined by using the bandpass task in {\sc cotter} on a pointed observation of 3C444. Imaging was carried out using standard {\sc miriad} routines. A robust weighting scheme of 0.0 was used to provide a good compromise between the high sensitivity provided by natural weighting while almost maintaining the synthesised beam of uniform weighting. The observing properties including sensitivity are summarised in table~.\ref{tab:obs}. Full details of the data reduction, imaging and flux calibration will be presented in \cite{Hindson2014}.
\newline

\section{\textbf  {Results}}\label{Sect:results}

\begin{figure*}
\centering
\includegraphics[width=1.0\textwidth]{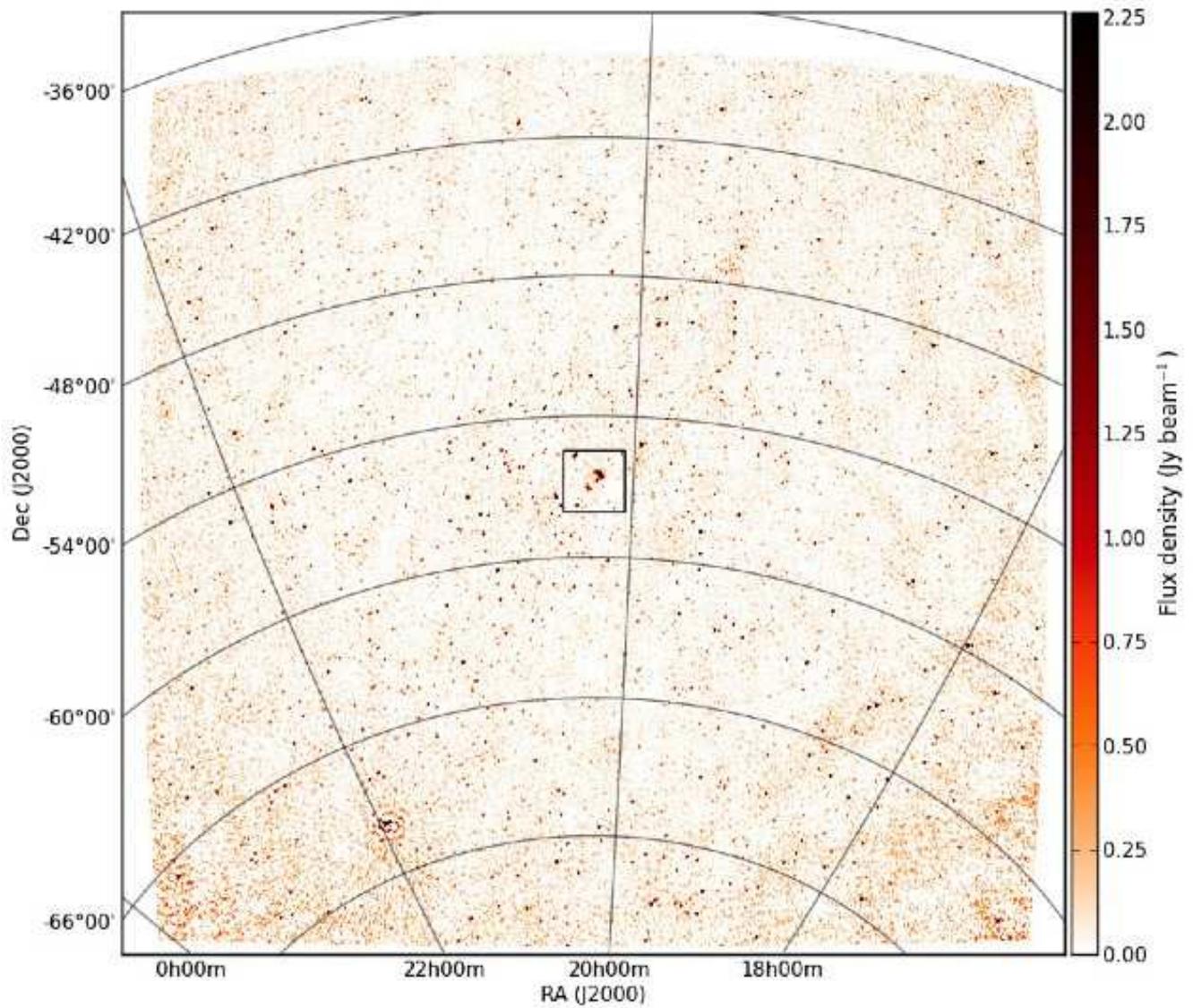} 
\caption{\textbf{Full FOV of the primary beam and flux corrected 120\,MHz image. The black box highlights the region we show in Fig~.\ref{im:zoom}.}
\label{im:radio}}
\end{figure*}

\begin{figure}
\centering
\includegraphics[width=0.45\textwidth]{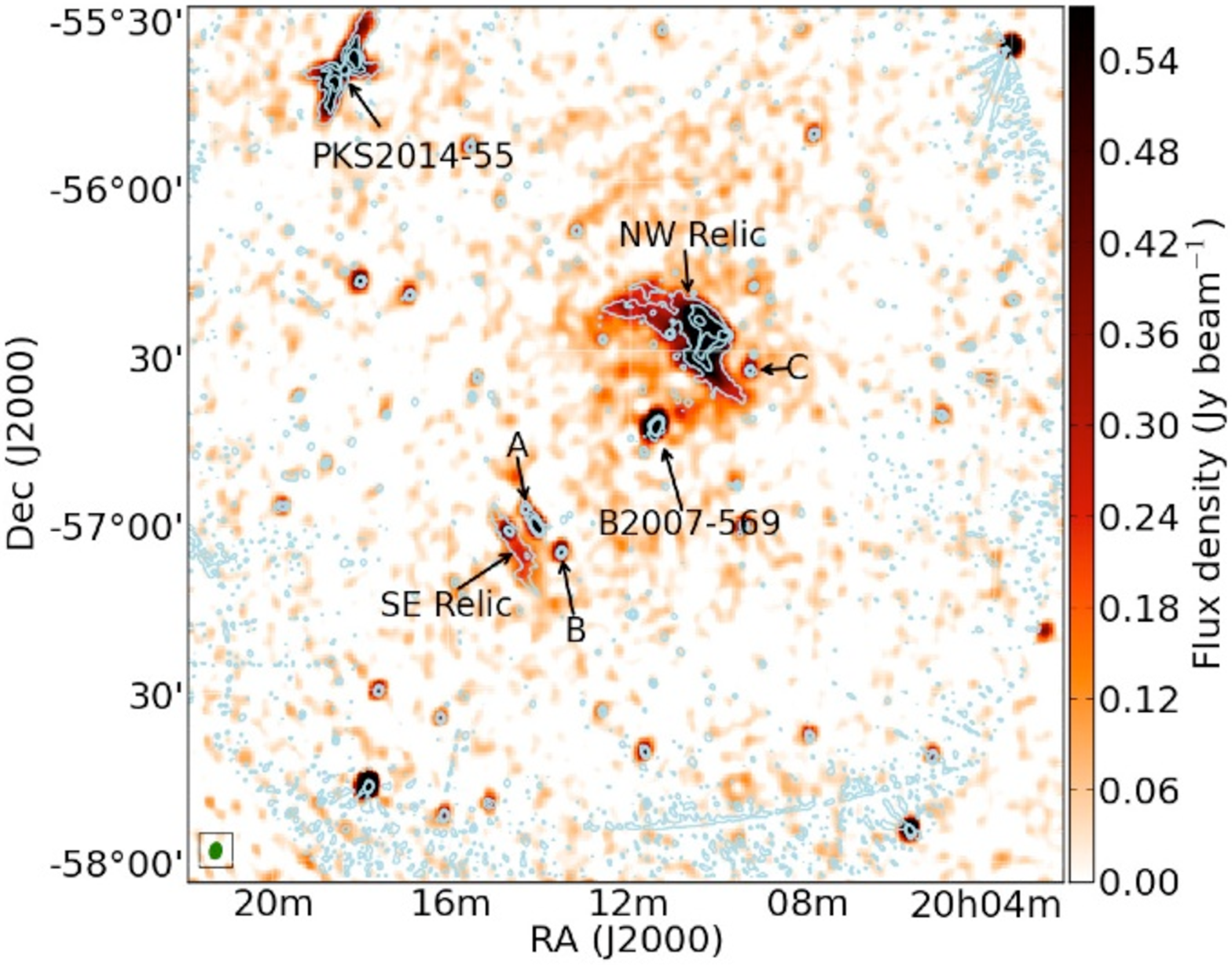} 
\includegraphics[width=0.45\textwidth]{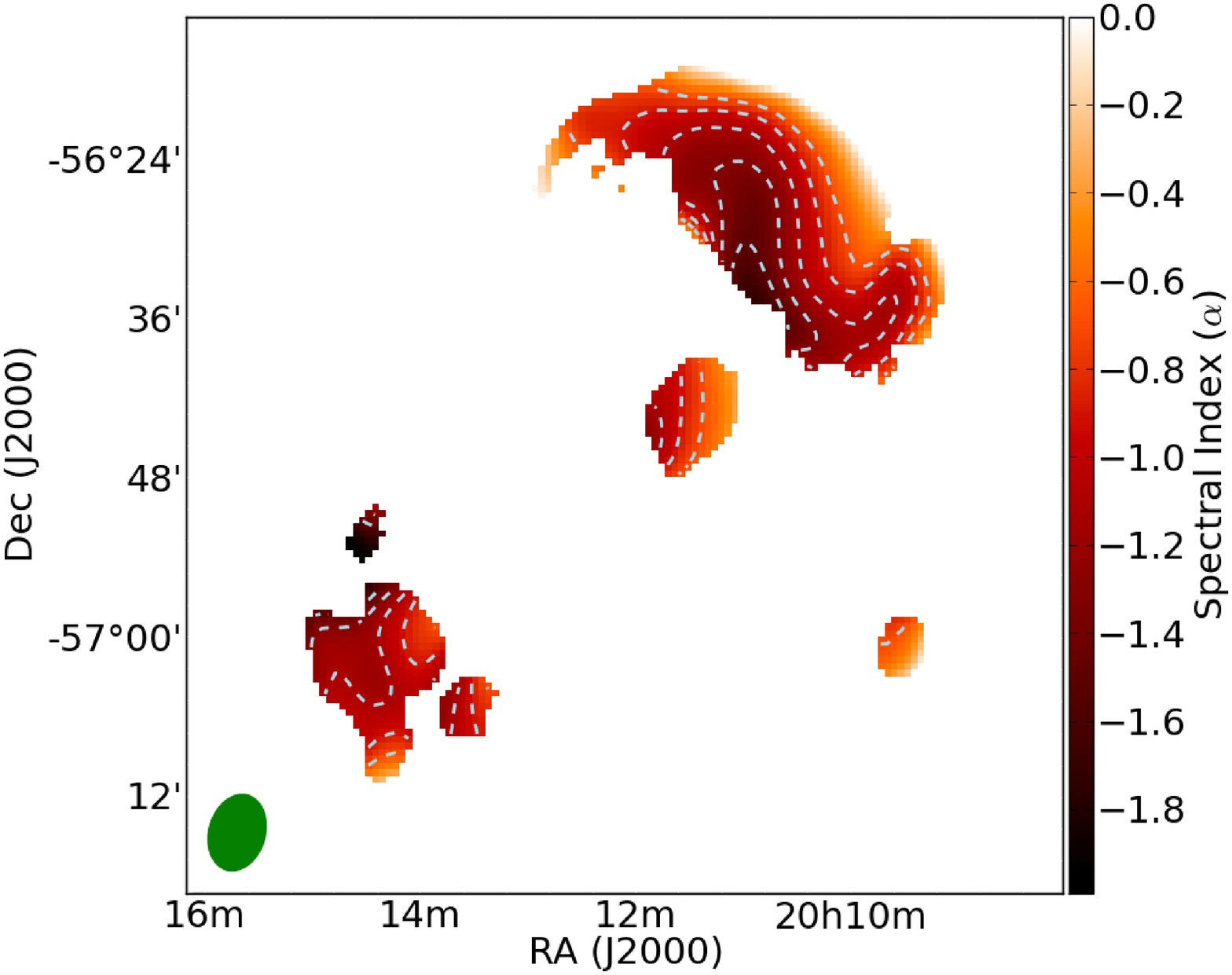} 
\caption{\textbf{Left panel: zoomed in image of A3667 at 226\,MHz. The synthesised beam is shown in the bottom left and contours show 843\,MHz data. Right panel: Spectral index map of A3667 from 120 to 843\,MHz.}}
\label{im:zoom}
\end{figure}

In Fig~.\ref{im:radio} we present the entire field of view (FOV) of our 120\,MHz image centred on A3667. This image highlights the incredibly wide FOV that makes the MWA an excellent survey instrument. We present a zoomed in image of the 226\,MHz band in Fig~.\ref{im:zoom} with contours taken from re-reduced Molonglo data \cite{Johnston2003} at 843\,MHz. 
\newline

These observations clearly detect the NW and SE relics on the periphery of the cluster. The peak brightness of the NW and SE relics is 4.6 and 1.1\,Jy\,beam$^{-1}$ at 149\,MHz respectively. At 120\,MHz we find that the SE relic becomes blended with sources labelled A and B. Towards the centre of the cluster we identify the bright head-tail radio galaxy B2007-569 which has a peak brightness of 5.4 Jy\,beam$^{-1}$ at 149\,MHz. We also highlight the radio galaxy PKS2014-55 which exhibits an interesting cross shaped morphology. 
\newline

\newpage

\section{\textbf  {Integrated Fluxes and Spectral Indices}}

We determine the integrated flux of the relics in A3667 by fitting apertures around the sources above 3$\sigma$. The integrated flux at 149\,MHz for the NW and SE relics is 28.1 and 2.4\,Jy, respectively. To explore the spectral properties of the two relics we subtract the flux of blended and contaminating point sources and fit the spectral index between our MWA bands and flux measurements taken from the literature using a power law. We find a spectral index between 120 and 1400\,MHz of $-0.9\pm0.1$ for both relics. 
\newline

\begin{wrapfigure}{r}{0.5\textwidth}
\includegraphics[width=0.45\textwidth,trim = 30 30 70 70]{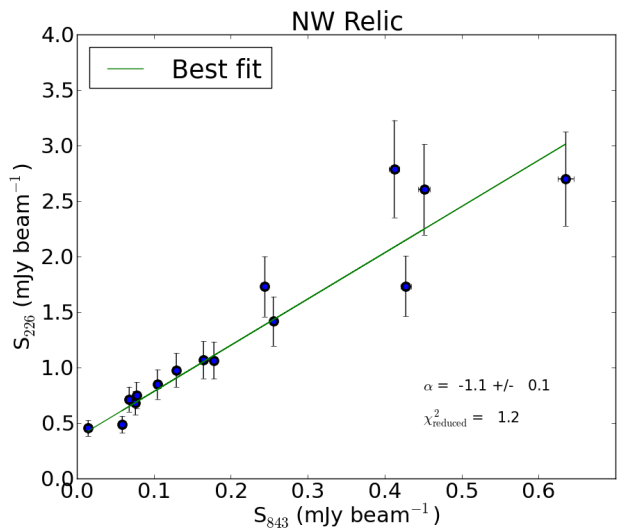} 
  \caption{\textbf{T-T plot of the NW relic between 226 to 843\,MHz with the best fit to the data, spectral index and chi-squared value.}}
  \label{im:TT}

\end{wrapfigure}

We generate spectral index maps of A3667 by first convolving our MWA images and Molonglo data with a common Gaussian kernel slightly larger than the beam at 120\,MHz. We then fit a power-law to each pixel above $3\sigma_{\rm rms}$. The resultant spectral index map can be seen in the right panel of Fig~.\ref{im:zoom}. The average spectral index of the NW relic in this map is $-1.0\pm0.2$ and varies from $-1.7$ towards to centre to $-0.4$ on the periphery. These results are consistent with the 3-point spectral index values between 843 MHz and 2.4 GHz \cite{Johnston2003}. The spectral index of the SE relic in this map is blended with sources A and B. However, using just the higher frequency observations will allow us to more accurately determine the spectral index of this relic.
\newline

We are able to generate a T-T plot of the large NW relic between 226 and 843\,MHz. We  convolve the 843\,MHz map with a Gaussian equal to the 226\,MHz beam and then regrid both maps so a single pixel corresponds to the size of the beam. Plotting the pixel flux at 843\,MHz against the corresponding flux at 226\,MHz in these maps and fitting a straight line results in an average spectral index for the NW relic of $-1.1\pm0.1$ (Fig~.\ref{im:TT}). This technique accounts for the possibility of constant offsets between maps due to missing short spacings or large-scale foreground contamination. Overall, all spectral index estimates for the NW relic are in good agreement with the widest frequency range value of $-0.9\pm0.1$. For a more detailed analysis and discussion of this data see \cite{Hindson2014}.

\section{\textbf {Acknowledgements}}
MJ-H acknowledges support from the Marsden Fund. This scientific work makes use of the Murchison Radio-astronomy Observatory, operated by CSIRO. We acknowledge the Wajarri Yamatji people as the traditional owners of the Observatory site. Support for the MWA comes from the U.S. National Science Foundation (grants AST-0457585, PHY-0835713, CAREER-0847753, and AST-0908884), the Australian Research Council (LIEF grants LE0775621 and LE0882938), the U.S. Air Force Office of Scientific Research (grant FA9550-0510247), and the Centre for All-sky Astrophysics (an Australian Research Council Centre of Excellence funded by grant CE110001020). Support is also provided by the Smithsonian Astrophysical Observatory, the MIT School of Science, the Raman Research Institute, the Australian National University, and the Victoria University of Wellington (via grant MED-E1799 from the New Zealand Ministry of Economic Development and an IBM Shared University Research Grant). The Australian Federal government provides additional support via the Commonwealth Scientific and Industrial Research Organisation (CSIRO), National Collaborative Research Infrastructure Strategy, Education Investment Fund, and the Australia India Strategic Research Fund, and Astronomy Australia Limited, under contract to Curtin University. We acknowledge the iVEC Petabyte Data Store, the Initiative in Innovative Computing and the CUDA Center for Excellence sponsored by NVIDIA at Harvard University, and the International Centre for Radio Astronomy Research (ICRAR), a Joint Venture of Curtin University and The University of Western Australia, funded by the Western Australian State government.

\end{document}